\documentclass[AMA,STIX1COL]{WileyNJD-v2}

\articletype{Article Type}%

\received{26 April 2016}
\revised{6 June 2016}
\accepted{6 June 2016}

\newcommand{\condind}{\mathrel{\text{\scalebox{1.07}{$\perp\mkern-10mu\perp$}}}}

\begin{document}

\title{An Inverse Probability Weighted Regression Method that Accounts for Right-censoring for Causal Inference with Multiple Treatments and a Binary Outcome.}

\author[1]{Youfei Yu}

\author[1]{Min Zhang*}

\author[1]{Bhramar Mukherjee}

\authormark{YOUFEI YU \textsc{et al}}

\address[1]{\orgdiv{Department of Biostatistics, School of
Public Health}, \orgname{University of Michigan, Ann Arbor}, \orgaddress{\state{Michigan}, \country{USA}}}



\corres{*Min Zhang, Department of Biostatistics, School of Public Health, University of Michigan, 1420 Washington Heights, Ann Arbor, MI 48109, USA. \email{mzhangst@umich.edu}}


\abstract[Summary]{Comparative effectiveness research often involves evaluating the differences in the risks of an event of interest between two or more treatments using observational data. Often, the post-treatment outcome of interest is whether the event happens within a pre-specified time window, which leads to a binary outcome. One source of bias for estimating the causal treatment effect is the presence of confounders, which are usually controlled using propensity score-based methods. An additional source of bias is right-censoring, which occurs when the information on the outcome of interest is not completely available due to dropout, study termination, or treatment switch before the event of interest. We propose an inverse probability weighted regression-based estimator that can simultaneously handle both confounding and right-censoring, calling the method CIPWR, with the letter C highlighting the censoring component. CIPWR estimates the average treatment effects by averaging the predicted outcomes obtained from a logistic regression model that is fitted using a weighted score function. The CIPWR estimator has a double robustness property such that estimation consistency can be achieved when either the model for the outcome or the models for both treatment and censoring are correctly specified. We establish the asymptotic properties of the CIPWR estimator for conducting inference, and compare its finite sample performance with that of several alternatives through simulation studies. The methods under comparison are applied to a cohort of prostate cancer patients from an insurance claims database for comparing the adverse effects of four candidate drugs for advanced stage prostate cancer.}

\keywords{causal inference, claims data, double robustness, observational studies, right censoring.}

\maketitle


\section{Introduction}
\label{s:intro}

Data from observational studies, in which treatments are usually not randomly assigned, have been increasingly used to evaluate comparative effectiveness of treatments in the real world. Without randomization, it is difficult to make causal interpretations concerning the effect of a treatment on an outcome of interest, as the estimation of the average treatment effect tends to be biased by the presence of confounders. A commonly used tool that controls for confounding is the propensity score, defined as the conditional probability of treatment given a set of potential confounders.\cite{rosenbaum1983} There is a large body of work on propensity score-based methods that adjust for the differences in baseline covariates among treatment groups, and readers can find some reviews of these methods by Stuart, \cite{stuart2010} Hu et al,\cite{hu2020} and Yu et al.\cite{yu2021}

It usually takes some period of follow-up time, which may or may not be the same for all subjects, to observe the post-treatment outcome of interest after the subjects enter the study. This type of outcome is sometimes referred to as time-lagged response. \cite{anstrom2001} One example is whether the event of interest happens within a pre-specified time window, which results in a binary outcome that cannot be ascertained if the subject has been censored prior to the end of the window. Censoring occurs when the information about the response is not completely available due to dropout, study termination, or treatment switch. Censoring together with confounding arises in many applications, such as insurance claims databases, where the participants are not randomly enrolled nor randomly assigned to treatments, and could potentially leave the insurance plan or be switched to another treatment prior to the occurrence of the event of interest. In this paper, we consider a study that evaluates the possible adverse effects of four candidate drugs prescribed for metastatic castration-resistant prostate cancer (mCRPC) using data from Optum Clinformatic Data Mart, a national private health insurance network. Some of these drugs are chemotherapy, while some others are hormone therapies, and the prior expectation is that the hormone therapies will lead to less acute adverse events. We separately examine two endpoints: emergency room (ER) visit and all-cause hospitalization post first-line therapy with one of these drugs. We intend to compare patients' risks of experiencing at least one event of interest (i.e., ER visit or hospitalization) within a 180-day, 270-day, and 360-day time window after treatment initiation. Therefore, the outcome we focus on is binary that is subject to right-censoring. 

Leaving out censored observations when conducting a comparative analysis can result in a biased effect estimate even after proper adjustment for confounding. Special techniques that adjust for both confounders and censoring are required to make valid inference on the causal effects. Anstrom and Tsiatis proposed an inverse probability weighted estimator of average treatment effect for the time-lagged response, which can be used to estimate the difference in risks of event occurrence.\cite{anstrom2001} To improve the robustness and statistical efficiency of the estimator of Anstrom and Tsiatis, Wang et al proposed an augmented inverse probability weighted estimator that makes use of the information about the outcome model for censored medical cost data.\cite{wang2016} The estimator of Wang et al can conveniently be extended to the case of a binary outcome by replacing the linear regression model for the outcome with a logistic regression model. Since one minus the survival function of the event of interest describes the risk of an event over the duration of follow-up, one can also use methods for time-to-event outcomes. These will include approaches that model the whole survival curve, and obtain the probability of surviving to the end of the pre-specified time window. For example, Zhang and Schaubel proposed an estimator for the cumulative hazard function, which incorporates the information from Cox models for the time-to-event outcome into the estimating equations.\cite{zhang2012} Their method was originally developed for estimation of restricted mean lifetimes, but can be easily adapted to estimate the average treatment effect on a possibly censored binary outcome, since one can think of the survival probability at a given time point as a fixed time `snapshot' of the whole survival curve. The augmented estimators such as those proposed by Wang et al\cite{wang2016} and Zhang and Schaubel\cite{zhang2012} are known to possess the double robustness property, such that the estimator remains consistent as long as either the model corresponding to the outcome, or the models corresponding to the weights in the estimating equations are correct. Another line of work in causal inference for censored data involves using pseudo-observations in replacement of the original outcomes that are possibly incompletely observed.\cite{andersen2017} For a binary outcome, the causal treatment effect is estimated by computing pseudo survival probability for each subject, followed by standard causal inference method for completely observed outcomes, such as inverse probability weighting or direct standardization. We leave the details of the aforementioned methods to Section \ref{s:comparative}. Among these approaches that address right-censoring, some assume conditional independence of the censoring and survival time given treatment only,\cite{anstrom2001,wang2016,andersen2017} while others require less restrictive conditions such that censoring and survival times are independent given treatment and baseline covariates.\cite{zhang2012}

We propose a method that directly models the binary outcome using logistic regression, with confounding and censoring properly accounted for by weighting. The risk of event occurrence (and therefore the average treatment effect) is estimated based on standardization, which averages the outcome predictions obtained from the logistic regression model across all subjects. The treatment assignment and censoring mechanism together can be viewed as a special case of coarsening, a process that prevents one from observing the desired data structure.\cite{tsiatis2006} We explain how our problem can be described using the concept of coarsening later in Section~\ref{s:CIPWR}. Coarsening is handled by applying inverse probability of not being coarsened as weights to the score function of the logistic regression model. We call the method inverse probability weighted regression-based estimator, CIPWR for short, with the letter C highlighting the censoring component. Specifically, three sets of working models are constructed, one for the treatment assignment, one for the treatment specific censoring distribution, and the other for the outcome of interest. We show that the CIPWR estimator is doubly robust in the sense that consistency of the estimator can be achieved if either the outcome or the coarsening mechanism is correctly modeled. As Zhang and Schaubel,\cite{zhang2012} this method makes the less restrictive assumption about censoring than Anstrom and Tsiatis,\cite{anstrom2001} Wang et al,\cite{wang2016} and Anderson et al.\cite{andersen2017} Unlike Zhang and Schaubel\cite{zhang2012} and Wang et al\cite{wang2016} that are based on the general approach of augmented inverse probability weighting, our method is a standardization method. Also, unlike Zhang and Schaubel\cite{zhang2012} that estimates the whole survival curve, this method targets the binary outcome of interest and may lead to improved efficiency in some situations.

The rest of the paper is organized as follows. In Section~\ref{s:notation}, we introduce the statistical framework and the notations used. In Section~\ref{s:CIPWR}, we describe our proposed method and establish its asymptotic properties. We compare the finite sample performance of the CIPWR estimator to that of several alternative approaches through simulation studies and results are presented in Section \ref{s:simulation}. The proposed method is then applied to the prostate cancer treatment comparison example from the insurance claims database in Section \ref{s:application}. Conclusions and discussions for future research are presented in Section \ref{s:discussion}.

\section{Notations and Assumptions}
\label{s:notation}
For individual $i$, where $i=1,\cdots,n$, let $\tilde{\boldsymbol{X}}_i$ be a set of baseline variables, and $Z_i$ be the treatment received. We assume that $Z_i$ is nominal with $J$ levels, i.e., $Z_i=j\in\{1,\cdots,J; J\ge 2\}$, and let $D_{ij}\equiv I(Z_i=j)$. Let $T_i$ denote the underlying lag time to the first event of interest, which will always be observed if there were no censoring. In this study, the outcome of interest, denoted by $Y_i$, is whether the event of interest occurs within a pre-specified time window $d$. By this definition, $Y_i=I(T_i<d)$. We adopt the counterfactual framework to formulate the problem of causal comparison \citep{rubin1974}. Each individual is associated with a set of potential outcomes $\{Y_i^{(1)},\cdots,Y_i^{(J)}\}$, where $Y_i^{(j)}=I\{T_i^{(j)}<d\}$ and $T_i^{(j)}$ is defined as the potentially observed time to the first event of interest had the patient received treatment $j$. Under the Stable Unit Treatment Value Assumption (SUTVA, defined later), only the outcome under the actual treatment received,  $Y_i=\sum_{j=1}^JD_{ij}Y_i^{(j)}$, can be observed. 

In practice, the time to event $T_i$ may not be completely observed due to right-censoring, in which case the outcome variable $Y_i$ is therefore subject to coarsening. Let $C_i$ denote the censoring time and $R_i=I\left\{C_i\ge\min(T_i,d)\right\}$. Then $Y_i$ is observed if the individual has not been censored before $d$, i.e., $R_i=1$. We further let $\Delta_i=I(T_i\le C_i)$ and $L_i=\min(T_i,C_i,d)$. Note that the outcomes $Y_i$ of those whose $T_i$ are censored ($\Delta_i=0$) are not necessarily missing at time $d$ ($R_i=0$).

Interest lies in estimating the average treatment effect $\tau\left(j,j'\right)=E\{Y^{\left(j'\right)}-Y^{(j)}\}$, which equals the risk difference $pr\{Y^{(j')}=1\} - pr\{Y^{(j)}=1\}$ for a binary outcome. We seek to estimate $E\{Y^{(j)}\}$ separately for $j=1,\cdots,J$. To connect the counterfactual framework to the observable data and establish a causal interpretation, we make the following assumptions.

\begin{enumerate}[I.]
    \item (\textit{Random sampling}) The individuals in the study are randomly sampled from the population.
    \item (\textit{Stable Unit Treatment Value Assumption, or SUTVA}) For any individual $i$, $i=1,\cdots,n$, if $Z_i=j$, then $Y_i=Y_i^{(j)}$, for all $j=1,\cdots,J$.
    \item (\textit{Unconfoundedness}) $\{Y_i^{(1)},\cdots, Y_i^{(J)}\}\condind Z_i|\tilde{\boldsymbol{X}}_i$. \label{assump:unconfound}
    \item (\textit{Overlap}) For all values of $j$ and $\tilde{\boldsymbol{x}}$, $0<\pi_j(\tilde{\boldsymbol{x}})<1$, where $\pi_j(\tilde{\boldsymbol{x}})=pr(Z_i=j|\tilde{\boldsymbol{x}})$.
    \item (\textit{Censoring at random}) $ C_i\condind\{T_i^{(1)},\cdots,T_i^{(J)}\}\Big|(Z_i, \tilde{\boldsymbol{X}}_i)$. \label{assump:censor}
\end{enumerate}

\section{Proposed Method: Inverse Probability Weighted Regression that Accounts for Right-Censoring}
\label{s:CIPWR}

We note that instead of directly evaluating $E\{Y_i^{(j)}\}$, it is theoretically more convenient to work with the survival function 
\begin{align*}
\mu_j\equiv E[I\{T_i^{(j)}\ge d\}] = 1-E\{Y_i^{(j)}\},
\end{align*}
and we let $\tilde{Y}_i^{(j)}=I\{T_i^{(j)}\ge d\}$. The counterfactual parameter $\mu_j$ can be represented using the observed data, $\mu_j=E_X[E\{\tilde{Y}_i^{(j)}|\tilde{\boldsymbol{X}}_i\}]=E_X[E\{\tilde{Y}_i|\tilde{\boldsymbol{X}}_i, Z_i=j\}]$, where the second equation follows from the unconfoundedness assumption \ref{assump:unconfound}. Had there been no right-censoring, $\mu_j$ could be estimated by averaging the predicted potential outcomes, $\hat{\mu}_j=n^{-1}\sum_{i=1}^n\hat{E}(\tilde{Y}_i|\tilde{\boldsymbol{X}}_i,Z_i=j)$, for $j=1,\cdots,J$, where $\hat{E}(\tilde{Y}_i|\tilde{\boldsymbol{X}}_i,Z_i=j)$ are usually fitted values in a parametric regression model. For a binary outcome, a logistic regression model is a popular choice to fit $\tilde{Y}$ in group $j$, specified as
\begin{align}
    \text{logit}\{E(\tilde{Y}_i|\tilde{\boldsymbol{X}}_i, Z_i=j)\}=\boldsymbol{X}_i^T\boldsymbol{\beta}_j, \label{eq:Q-model}
\end{align}
where $\boldsymbol{X}_i$ is a vector-valued function of $\tilde{\boldsymbol{X}}_i$ with an intercept and possibly interactions and nonlinear terms. For notational convenience, we define $m_{ij}(\boldsymbol{\beta}_j)=\text{expit}(\boldsymbol{X}_i^T\boldsymbol{\beta}_j)$. When the outcome $Y$ (and therefore $\tilde{Y}$) is completely observed for all individuals in the sample, $\boldsymbol{\beta}_j$ is commonly estimated by solving the score equations
\begin{align}
    \sum_{i=1}^n\boldsymbol{X}_i\{\tilde{Y}_i-m_{ij}(\boldsymbol{\beta}_j)\}=0, \label{eq:score_complete}
\end{align}
the solution of which is the maximum likelihood estimator.

From the missing data perspective, the potential outcome $Y_i^{(j)}=I\{T_i^{(j)}<d\}$ would be missing at baseline ($t=0$) if individual $i$ were assigned to the treatment other than $j$. When censoring comes into play, $Y_i^{(j)}$ is subject to missingness at any time $0<t<d$ because of censoring. In this case, we consider the more general notion of coarsening of data \cite{tsiatis2006,heitjan1991,gill1997}, which describes the case where one only gets to observe a many-to-one function of the full data for some of the individuals in the sample, and different many-to-one functions are allowed for different individuals. In the context of estimating $\mu_j$, the full data one would like to observe for individual $i$ is $\{Y_i^{(j)},\tilde{\boldsymbol{X}}_i\}$. When $D_{ij}=0$, $T_i^{(j)}$ (and therefore $Y_i^{(j)}$) is completely missing, and we only observe $\tilde{\boldsymbol{X}}_i$. When $D_{ij}=1$ and $C_i=t<\min(T_i^{(j)}, d)$, we observe $\{I(T_i^{(j)}>t), \tilde{\boldsymbol{X}}_i\}$, where $t<d$. When $D_{ij}=1$ and $C_i=t\ge\min(T_i^{(j)}, d)$, there was no coarsening at all, and we observe the full data $\{Y_i^{(j)}, \tilde{\boldsymbol{X}}_i\}$. In summary, there are two layers of missingness in our setting, one due to treatment assignment and the other due to censoring. Therefore, we can inversely weight the unbiased estimating equations (\ref{eq:score_complete}) by the probability of not being coarsened to make inference about the target population where no subjects were coarsened. The weighted estimating equations for $\boldsymbol{\beta}_j$ is given by
\begin{align}
\sum_{i=1}^n\frac{D_{ij}R_i\boldsymbol{X}_i \{\tilde{Y}_i-m_{ij}(\boldsymbol{\beta}_j)\}}{\pi_{ij}(\boldsymbol{\alpha})\exp\{{-\Lambda_{ij}(L_i)}\}} = 0, \label{eq:weightedscore}
\end{align}
where $\pi_{ij}(\boldsymbol{\alpha})=pr(Z_i=j|\tilde{\boldsymbol{X}}_i)$ is the propensity score for treatment $j$, and $\Lambda_{ij}(t)$ is the cumulative hazard function of $C_i$ at $t$ for treatment $j$. We denote the solution to (\ref{eq:weightedscore}) by $\hat{\boldsymbol{\beta}}_j$. The total weights that account for the coarsening mechanism consist of two weighting components. The first is the propensity of being assigned to treatment $j$, and the second is (informally) the conditional probability of not being censored. Therefore, equation~(\ref{eq:weightedscore}) can be regarded as weighting the score functions (\ref{eq:score_complete}) by the inverse probability of observing the complete cases. A regression-based estimator for $\mu_j$ indicated by (\ref{eq:Q-model}) is
\begin{align*}
    \hat{\mu}_j=n^{-1}\sum_{i=1}^n\text{expit}(\boldsymbol{X}_i^T\hat{\boldsymbol{\beta}}_j)=n^{-1}\sum_{i=1}^n m_{ij}(\hat{\boldsymbol{\beta}}_j),
\end{align*}
and we call it inverse probability weighted regression-based estimator that accounts for right-censoring (CIPWR).

In the literature, an outcome that is subject to censoring is sometimes handled using survival models, such as Cox regression model.\cite{zhang2012,cox1972} CIPWR, on the other hand, directly models the binary outcome of interest using the logistic regression, which is relatively more intuitive and straightforward to implement for empirical researchers. Our proposed estimator can be implemented using standard statistical software, such as the \textit{glm} function in R with the \textit{weights} argument being specified.

In practice, $\pi_{ij}(\boldsymbol{\alpha})$ and $\Lambda_{ij}(t)$ in (\ref{eq:weightedscore}) are usually unknown and need to be estimated from the data. We build working models for these two nuisance components. Let $\boldsymbol{V}_i$ and $\boldsymbol{W}_i$ be vector-valued functions of $\tilde{\boldsymbol{X}}_i$, which are allowed to be different from $\boldsymbol{X}_i$. We assume that the treatment assignment mechanism is governed by a multinomial logistic regression model
\begin{align*}
    \log\frac{pr(Z_i=j|\boldsymbol{V}_i)}{pr(Z_i=J|\boldsymbol{V}_i)} = \boldsymbol{V}_i^T\boldsymbol{\alpha}_j,\hspace{2mm} j=1,\cdots,J-1,
\end{align*}
where $J$ is the reference treatment level. Let $\boldsymbol{\alpha}=\left(\boldsymbol{\alpha}_1,\cdots,\boldsymbol{\alpha}_{J-1}\right)^T$, and its estimated value $\hat{\boldsymbol{\alpha}}$ can be obtained through maximum likelihood estimation. With respect to censoring, for each treatment $j=1,\cdots,J$, we assume a Cox proportional hazards model, specified as
\begin{align*}
    \lambda_{ij}(t|\boldsymbol{W}_i,\boldsymbol{\gamma}_j)=\lambda_{0j}(t)\exp(\boldsymbol{W}_i^T\boldsymbol{\gamma}_j),
\end{align*}
where $\lambda_{0j}(t)$ is an unspecified treatment-specific baseline hazard function of $C$. The estimates for $\boldsymbol{\gamma}_j$ and $\Lambda_{0j}(t)=\int_0^t\lambda_{0j}(s)ds$ can be determined by the maximum partial likelihood estimator, $\hat{\boldsymbol{\gamma}}_j$, and the Breslow estimator, $\hat{\Lambda}_{0j}(t)$, respectively. Then the probability of remaining uncensored at $t$ for individual $i$ is given by $\exp\{-\hat{\Lambda}_{0j}(t)\exp(\boldsymbol{W}_i^T\boldsymbol{\hat{\gamma}}_j)\}$. In a typical survival study, one only gets to observe the minimum of $C$ and $T$, which is usually referred to as \textit{observation time}, and the observed data can be represented as $\{\Delta_i, \min(T_i,C_i)\}$. However, for our data example, time to treatment switch or the end of insurance coverage can always be identified from the claims data, regardless of whether the event of interest happens or not. In this case, censoring time $C$ is always available for all subjects, and $\Delta_i=0$ for all $i, i=1,\cdots,n$. Therefore, one can alternatively estimate the probability of remaining uncensored by replacing $\{\Delta_i, \min(T_i,C_i)\}$ with $\{0, C_i\}$.

\subsection{Consistency and Double Robustness}
\label{s:consistency}
Under suitable regularity conditions, $\hat{\boldsymbol{\alpha}}$, $\hat{\boldsymbol{\gamma}}_j$, and $\hat{\Lambda}_{0j}$ converge in probability to well-defined limits, denoted by $\boldsymbol{\alpha}^\ast$, $\boldsymbol{\gamma}_j^\ast$, and $\boldsymbol{\Lambda}_{0j}^\ast$, respectively, which can be different from their corresponding true values $\boldsymbol{\alpha}^0$, $\boldsymbol{\gamma}_j^0$, and $\boldsymbol{\Lambda}_{0j}^0$ \citep{casella2002, lin1989}. We denote the true values for $\boldsymbol{\beta}_j$ and $\mu_j$ by $\boldsymbol{\beta}_j^0$ and $\mu_j^0$, respectively. For notational convenience, we also define $\Lambda_{ij}^\ast(t)=\Lambda_{0j}^\ast(t)\exp(\boldsymbol{W}_i^T\boldsymbol{\gamma}_j^\ast)$ and $\Lambda_{ij}^0(t)=\Lambda_{0j}^0(t)\exp(\boldsymbol{W}_i^T\boldsymbol{\gamma}_j^0)$. 


We first show that $\hat{\mu}_j=n^{-1}\sum_{i=1}^nm_{ij}(\hat{\boldsymbol{\beta}}_j)$, a function of $\hat{\boldsymbol{\beta}}_j$, is consistent when the outcome model for treatment $j$ is correct. Using the theory of M-estimator, the consistency of $\hat{\boldsymbol{\beta}}_j$ can be established by showing that the estimating function is unbiased \citep{Boos2013}, that is, 

\begin{subequations}
 \begin{align}
    0=&E\left[\frac{D_{ij}R_i\boldsymbol{X}_i\{\tilde{Y}_i-m_{ij}(\boldsymbol{\beta}_j^0)\}}{\pi_{ij}(\boldsymbol{\alpha}^\ast)\exp\{-\Lambda^\ast_{ij}(L_i)\}}\right] \nonumber \\ 
    =&E\left[\frac{D_{ij}R_i\boldsymbol{X}_i\tilde{Y}_i}{\pi_{ij}(\boldsymbol{\alpha}^\ast)\exp\{-\Lambda^\ast_{ij}(L_i)\}}\right] \label{eq:ef1} \\
    &- E\left[\frac{D_{ij}R_i\boldsymbol{X}_i m_{ij}(\boldsymbol{\beta}_j^0)}{\pi_{ij}(\boldsymbol{\alpha}^\ast)\exp\{-\Lambda^\ast_{ij}(L_i)\}}\right]. \label{eq:ef2}
\end{align}
\end{subequations}

Applying the law of iterated expectation and using the Assumption \ref{assump:censor},
\begin{align*}
    (\ref{eq:ef1}) &=E\left\{E\left[\frac{D_{ij}R_i\boldsymbol{X}_i\tilde{Y}_i}{\pi_{ij}(\boldsymbol{\alpha}^\ast)\exp\{-\Lambda^\ast_{ij}(L_i)\}}\middle| \boldsymbol{X}_i,Z_i=j\right] \right\} \\
    &=E\left[\frac{D_{ij}\boldsymbol{X}_iE\left\{I(C_i>d)\middle|\boldsymbol{X}_i,Z_i=j\right\}}{\pi_{ij}(\boldsymbol{\alpha}^\ast)\exp\{-\Lambda^\ast_{ij}(d)\}} E(\tilde{Y}_i|\boldsymbol{X}_i,Z_i=j) \right],
    \end{align*}
where the second equation is derived from the formula $R_i\tilde{Y}_i=I\{C_i>\min(T_i,d)\}\tilde{Y}_i=I\{C_i>d\}\tilde{Y}_i$. Since when $T_i>d$, $R_i/\exp\{-\Lambda_{ij}^\ast(L_i)\}=I(C_i>d)/\exp\{-\Lambda_{ij}^\ast(d)\}$, using similar techniques,
\begin{align*}
    (\ref{eq:ef2}) &= E\left\{E\left[\frac{D^{(j)}R_i\boldsymbol{X}_i m_{ij}(\boldsymbol{\beta}_j^0)}{\pi_{ij}(\boldsymbol{\alpha}^\ast)\exp\{-\Lambda^\ast_{ij}(L_i)\}}\middle|\boldsymbol{X}_i,Z_i=j,T_i>d\right]\right\} \\
    &=E\left\{\frac{D_{ij}\boldsymbol{X}_iE\left[I(C_i>d)\middle|\boldsymbol{X}_i,Z_i=j\right]}{\pi_{ij}(\boldsymbol{\alpha}^\ast)\exp\{-\Lambda^\ast_{ij}(d)\}}m_{ij}(\boldsymbol{\beta}_j^0) \right\}.
\end{align*}

Since $E(\tilde{Y}_i|\boldsymbol{X}_i,Z_i=j)=m_{ij}(\boldsymbol{\beta}_j^0)$ when the outcome model is correctly specified, the estimating function is shown to be unbiased, which implies that $\hat{\boldsymbol{\beta}}_j$ obtained by solving (\ref{eq:weightedscore}) converges in probability to the truth $\boldsymbol{\beta}_j^0$. Therefore, $\hat{\mu}_j = n^{-1}\sum_{i=1}^n m_{ij}(\hat{\boldsymbol{\beta}}_j) \overset{p}{\to} E\left\{m_{ij}\left(\boldsymbol{\beta}_j^0\right)\right\}=\mu_j^0$.

We then show the consistency of $\hat{\mu}_j$ when the coarsening mechanisms (i.e., treatment and censoring models) are correctly specified, in which case $\pi_{ij}(\hat{\boldsymbol{\alpha}})\overset{p}{\to}\pi_{ij}(\boldsymbol{\alpha}^0)$ and $\hat{\Lambda}_{ij}(t)\overset{p}{\to}\Lambda_{ij}^0(t)$. Under suitable regularity conditions, $\hat{\boldsymbol{\beta}}_j\overset{p}{\to}\boldsymbol{\beta}_j^\ast$, where $\boldsymbol{\beta}_j^\ast$ is a well-defined limit, and then
\begin{align*}
    \hat{\mu}_j = n^{-1}\sum_{i=1}^nm_{ij}\left(\hat{\boldsymbol{\beta}}_j\right) \overset{p}{\to} E\left\{m_{ij}\left(\boldsymbol{\beta}_j^\ast\right)\right\}.
\end{align*}

We consider the intercept term in $\boldsymbol{X}_i$ and rearrange equation (\ref{eq:weightedscore}),
\begin{align}
    n^{-1}\sum_{i=1}^n\frac{D_{ij}R_i\tilde{Y}_i}{\pi_{ij}(\hat{\boldsymbol{\alpha}})\exp\{-\hat{\Lambda}_{ij}(L_i)\}} = n^{-1}\sum_{i=1}^n\frac{D_{ij}R_i m_{ij}(\hat{\boldsymbol{\beta}}_j)}{\pi_{ij}(\hat{\boldsymbol{\alpha}})\exp\{-\hat{\Lambda}_{ij}(L_i)\}}.
    \label{eq:weightedscore_intercept}
    \end{align}
The left-hand side of (\ref{eq:weightedscore_intercept}) converges in probability to $\mu_j^0$, because
\begin{align}
        n^{-1}\sum_{i=1}^n\frac{D_{ij}R_i\tilde{Y}_i}{\pi_{ij}(\hat{\boldsymbol{\alpha}})\exp\{-\hat{\Lambda}_{ij}(L_i)\}}&\xrightarrow{\text{ }p\text{ }}E\left[\frac{D_{ij}R_i\tilde{Y}_i}{\pi_{ij}(\boldsymbol{\alpha}^0)\exp\{-\Lambda_{ij}^0(L_i)\}}\right]  \nonumber\\
        &=E\left[\frac{D_{ij}E\{I(C_i>d)\tilde{Y}_i^{(j)}|\tilde{\boldsymbol{X}}_i, Z_i=j\}}{\pi_{ij}(\boldsymbol{\alpha}^0)\exp\{-\Lambda_{ij}^0(d)\}}\right] \nonumber \\
        &= E\left[\frac{D_{ij}E\{I(C_i>d)|\tilde{\boldsymbol{X}}_i, Z_i=j\}E\{\tilde{Y}_i^{(j)}|\tilde{\boldsymbol{X}}_i, Z_i=j\}}{\pi_{ij}(\boldsymbol{\alpha}^0)\exp\{-\Lambda_{ij}^0(d)\}}\right] \label{eq:censoring_assump}
    \end{align}
where (\ref{eq:censoring_assump}) follows from Assumption~\ref{assump:censor}. With correct specification of the treatment and censoring models, $E\{I(C_i>d)|\tilde{\boldsymbol{X}}_i, Z_i=j\}=\exp\{-\Lambda_{ij}^0(d)\}$ and $E(D_{ij}|\tilde{\boldsymbol{X}}_i)=\pi_{ij}(\boldsymbol{\alpha}_0)$, and therefore (\ref{eq:censoring_assump}) can be reduced to $E[E\{\tilde{Y}_i^{(j)}|\tilde{\boldsymbol{X}}_i, Z_i=j\}]$, which is equivalent to $\mu_j^0$.

Using similar techniques, one can show that the right-hand side of (\ref{eq:weightedscore_intercept}) converges in probability to $E\left\{m_{ij}(\boldsymbol{\beta}_j^\ast)\right\}$, since
\begin{align*}
    n^{-1}\sum_{i=1}^n\frac{D_{ij}R_i m_{ij}(\hat{\boldsymbol{\beta}}_j)}{\pi_{ij}(\hat{\boldsymbol{\alpha}})\exp\{-\hat{\Lambda}_{ij}(L_i)\}} &\xrightarrow{\text{ }p\text{ }} E\left[\frac{D_{ij}R_i m_{ij}(\boldsymbol{\beta}_j^\ast)}{\pi_{ij}(\boldsymbol{\alpha}^0)\exp\{-\Lambda_{ij}^0(L_i)\}}\right] \\ &=E\left[\frac{D_{ij}m_{ij}(\boldsymbol{\beta}_j^\ast)E\{I(C_i>d)|\boldsymbol{X}_i,Z_i=j,T_i>d\}}{\pi_{ij}(\boldsymbol{\alpha}^0)\exp\{-\Lambda_{ij}^0(d)\}}\right] \\
    &=E\left\{m_{ij}(\boldsymbol{\beta}_j^\ast)\right\}.
\end{align*}

It follows that $\mu_j^0=E\left\{m_{ij}(\boldsymbol{\beta}_j^\ast)\right\}$, and $\hat{\mu}_j\overset{p}{\to}\mu_j^0$ when the treatment and censoring models are correctly specified. Note that when the stronger independence assumption ($C\condind T|Z$) holds for survival and censoring times, only the treatment model is required to be correct. We have shown that the proposed estimator exhibits the so-called double robustness property.

\subsection{Asymptotic Properties}
\label{s:asymptotic}
In this section, we establish the asymptotic properties of our proposed estimator $\hat{\mu}_j$. For $j=1,\cdots, J$, through a Taylor series expansion of $\hat{\mu}_j=n^{-1}\sum_{i=1}^nm_{ij}(\hat{\boldsymbol{\beta}}_j)$ about $\boldsymbol{\beta}_j^\ast$,
\begin{align}
    n^{1/2}(\hat{\mu}_j-\mu_j^0)= n^{-1/2}\sum_{i=1}^n\left\{ m_{ij}(\boldsymbol{\beta}^\ast_j) -\mu_j^0 \right\} + \boldsymbol{A}_j(\boldsymbol{\beta}_j^\ast)n^{1/2}(\hat{\boldsymbol{\beta}}_j-\boldsymbol{\beta}_j^\ast) + o_p(1),
    \label{eq:asymp_mu}
\end{align}
where $\boldsymbol{A}_j(\boldsymbol{\beta}_j^\ast) = E\left[\boldsymbol{X}_i^Tm_{ij}(\boldsymbol{\beta}_j^\ast)\{1-m_{ij}(\boldsymbol{\beta}_j^\ast)\}\right]$.

Equation (\ref{eq:asymp_mu}) indicates that to characterize the asymptotic distribution of $n^{1/2}(\hat{\mu}_j-\mu_j^0)$, one first needs to identify the asymptotic distribution of $n^{1/2}(\hat{\boldsymbol{\beta}}_j-\boldsymbol{\beta}_j^\ast)$, which further depends on the asymptotic results for the parameters of the treatment and censoring models. Under some suitable regularity conditions, $\hat{\boldsymbol{\alpha}}_l\overset{p}{\to}\boldsymbol{\alpha}_l^\ast$ for $l=1,\cdots,J-1$, and the estimator of the treatment model parameter is asymptotically normal with 
\begin{align}
    n^{1/2}(\hat{\boldsymbol{\alpha}}_l-\boldsymbol{\alpha}_l^\ast)=\boldsymbol{H}_l^{-1}(\boldsymbol{\alpha}^\ast)n^{-1/2}\sum_{i=1}^n\boldsymbol{V}_i\left\{D_{il}-\pi_{il}(\boldsymbol{\alpha}^\ast)\right\} + o_p(1), \label{eq:asymp_alpha}
\end{align}
where $\boldsymbol{H}_l(\boldsymbol{\alpha}^\ast) = E\left[\sum_{i=1}^n\boldsymbol{V}_i\boldsymbol{V}_i^T\pi_{il}(\boldsymbol{\alpha}^\ast)\left\{1-\pi_{il}(\boldsymbol{\alpha}^\ast)\right\}\right]$ with $\boldsymbol{\alpha}^\ast=(\boldsymbol{\alpha}_1^\ast, \cdots, \boldsymbol{\alpha}_{J-1}^\ast)^T$.

For the asymptotic distributions of the estimators $\hat{\gamma}_j$ and $\hat{\Lambda}_{ij}$, we define the relevant notations $\boldsymbol{s}_j^{(q)}(t;\boldsymbol{\gamma}_j)$ for $q=0,1,2$, $\overline{\boldsymbol{w}}_j(t;\boldsymbol{\gamma}_j)$, $d\Lambda_{0j}^\ast(t)$, and $dM_{ij}^\ast(t)$ in section A in the Supplementary Materials available at \textit{Biostatistics} online. We further denote the counting process by $N_{ij}(t)=D_{ij}I\{\min(T_i,C_i)\le t, \Delta_i=1\}$ and the at-risk process by $Y_{ij}(t)=D_{ij}I\{\min(T_i,C_i)\ge t\}$. Let $\delta$ be the time point that satisfies $P\{\min(T_i,C_i)\ge \delta\}>0$ for $i=1,\cdots,n$, which is practically set to the maximum observation time. Lin and Wei\cite{lin1989} showed that under some regularity conditions, $\hat{\boldsymbol{\gamma}}_j\overset{p}{\to}\boldsymbol{\gamma}_j^\ast$, and $n^{1/2}(\hat{\boldsymbol{\gamma}}_j-\boldsymbol{\gamma}_j^\ast)$ converges in distribution to a normal distribution
\begin{align}
    n^{1/2}(\hat{\boldsymbol{\gamma}}_j-\boldsymbol{\gamma}_j^\ast) = \boldsymbol{\Omega}^{-1}_j(\boldsymbol{\gamma}_j^\ast)n^{-1/2}\sum_{i=1}^n\boldsymbol{U}_{ij}(\boldsymbol{\gamma}_j^\ast) + o_p(1), \label{eq:asymp_gamma}
\end{align}
where $\boldsymbol{\Omega}_j(\boldsymbol{\gamma}_j^\ast)=\int_0^\delta\left\{\frac{\boldsymbol{s}_j^{(2)}(t;\boldsymbol{\gamma}_j^\ast)}{s_j^{(0)}(t;\boldsymbol{\gamma}_j^\ast)}-\overline{\boldsymbol{w}}_j(t;\boldsymbol{\gamma}_j)^{\otimes 2}\right\}E\{Y_{ij}(t)\lambda_{ij}(t)\}dt$ and $\boldsymbol{U}_{ij}(\boldsymbol{\gamma}_j^\ast)=\int_0^\delta\{\boldsymbol{W}_i-\overline{\boldsymbol{w}}(t;\boldsymbol{\gamma}_j^\ast)\}dM_{ij}^\ast(t)$. Using (\ref{eq:asymp_gamma}), one can show that
\begin{multline}
    n^{1/2}\{\hat{\Lambda}_{ij}(t)-\Lambda_{ij}^\ast(t)\}=\boldsymbol{K}_{ij}^T(t;\boldsymbol{\gamma}_j^\ast)\boldsymbol{\Omega}^{-1}(\boldsymbol{\gamma}_j^\ast)n^{-1/2}\sum_{i=1}^n\boldsymbol{U}_{ij}(\boldsymbol{\gamma}_j^\ast) \\
    +\exp(\boldsymbol{W}_i^T\boldsymbol{\gamma}_j^\ast)n^{-1/2}\sum_{i=1}^n\int_0^t\frac{dM_{ij}^\ast(u)}{s^{(0)}(u;\boldsymbol{\gamma}_j^\ast)} + o_p(1), 
    \label{eq:asymp_Lambda}
\end{multline}
where $\boldsymbol{K}_{ij}(t;\boldsymbol{\gamma}_j^\ast)=\int_0^t\{\boldsymbol{W}_i-\overline{\boldsymbol{w}}_j(t;\boldsymbol{\gamma}_j^\ast)\}d\Lambda_{ij}^\ast(u)$.

By a sequence of Taylor series expansion of $n^{-1}\sum_{i=1}^n\frac{D_{ij}R_i\boldsymbol{X}_i\{\tilde{Y}_i-m_{ij}(\hat{\boldsymbol{\beta}}_j)\}}{\pi_{ij}(\hat{\boldsymbol{\alpha}}) \exp\{-\hat{\Lambda}_{ij}(L_i)\}}$ (see Section A in the Supplementary Materials) and combining the results of (\ref{eq:asymp_alpha}), (\ref{eq:asymp_gamma}), and (\ref{eq:asymp_Lambda}), it follows that
\begin{multline}
    n^{1/2}(\hat{\boldsymbol{\beta}}_j-\boldsymbol{\beta}_j^\ast)=\boldsymbol{B}_j^{-1}(\boldsymbol{\beta}_j^\ast,\boldsymbol{\alpha}^\ast,\Lambda_{ij}^\ast)n^{-1/2}\sum_{i=1}^n\frac{D_{ij}R_i\boldsymbol{X}_i\{\tilde{Y}_i-m_{ij}(\boldsymbol{\beta}_j^\ast)\}}{\pi_{ij}(\boldsymbol{\alpha}^\ast) \exp\{-\Lambda_{ij}^\ast(L_i)\}} \\
    +\boldsymbol{B}_j^{-1}(\boldsymbol{\beta}_j^\ast,\boldsymbol{\alpha}^\ast,\Lambda_{ij}^\ast)\sum_{l=1}^{J-1}\left[\boldsymbol{F}_{jl}\left(\boldsymbol{\beta}_j^\ast,\boldsymbol{\alpha}^\ast,\Lambda_{ij}^\ast\right)\boldsymbol{H}_l^{-1}(\boldsymbol{\alpha}^\ast)n^{-1/2}\sum_{i=1}^n\boldsymbol{V}_i\{D_{il}-\pi_{il}(\boldsymbol{\alpha}^\ast)\}\right] \\
    +\boldsymbol{B}_j^{-1}(\boldsymbol{\beta}_j^\ast,\boldsymbol{\alpha}^\ast,\Lambda_{ij}^\ast)\boldsymbol{P}_j(\boldsymbol{\beta}_j^\ast,\boldsymbol{\alpha}^\ast,\Lambda_{ij}^\ast)\boldsymbol{\Omega}_j^{-1}(\boldsymbol{\gamma}_j^\ast)n^{-1/2}\sum_{i=1}^n\boldsymbol{U}_{ij}(\boldsymbol{\gamma}_j^\ast) \\
    +\boldsymbol{B}_j^{-1}(\boldsymbol{\beta}_j^\ast,\boldsymbol{\alpha}^\ast,\Lambda_{ij}^\ast)\boldsymbol{Q}_j(\boldsymbol{\beta}_j^\ast,\boldsymbol{\alpha},\Lambda_{ij}^\ast)n^{-1/2}\sum_{i=1}^n\int_0^t\frac{dM_{ij}^\ast(u)}{s^{(0)}(u;\boldsymbol{\gamma}_j^\ast)}+o_p(1).
    \label{eq:asymp_beta}
\end{multline}
where $\boldsymbol{B}_j$, $\boldsymbol{F}_{jl}$, $\boldsymbol{P}_j$, and $\boldsymbol{Q}_j$ are defined in section A in the Supplementary Materials available at \textit{Biostatistics} online.

Plugging (\ref{eq:asymp_beta}) into (\ref{eq:asymp_mu}), we can represent $n^{1/2}(\hat{\mu}_j-\mu_j)$ as $n^{-1/2}\sum_{i=1}^n\psi_{ij}+o_p(1)$, where
\begin{align*}
    \psi_{ij} =& m_{ij}(\boldsymbol{\beta}^\ast_j)-\mu_j+\boldsymbol{A}_j(\boldsymbol{\beta}_j^\ast)\boldsymbol{B}_j^{-1}(\boldsymbol{\beta}_j^\ast,\boldsymbol{\alpha}^\ast,\Lambda_{ij}^\ast)\frac{D_{ij}R_i\boldsymbol{X}_i\{\tilde{Y}_i-m_{ij}(\boldsymbol{\beta}_j^\ast)\}}{\pi_{ij}(\boldsymbol{\alpha}^\ast) \exp\{-\Lambda_{ij}^\ast(L_i)\}} \\
    &+\boldsymbol{A}_j(\boldsymbol{\beta}_j^\ast)\boldsymbol{B}_j^{-1}(\boldsymbol{\beta}_j^\ast,\boldsymbol{\alpha}^\ast,\Lambda_{ij}^\ast)\sum_{l=1}^{J-1}\boldsymbol{F}_{jl}\left(\boldsymbol{\beta}_j^\ast,\boldsymbol{\alpha}^\ast,\Lambda_{ij}^\ast\right)\boldsymbol{H}_l^{-1}(\boldsymbol{\alpha}^\ast)\boldsymbol{V}_i\{D_{il}-\pi_{il}(\boldsymbol{\alpha}^\ast)\} \\
    &+\boldsymbol{A}_j(\boldsymbol{\beta}_j^\ast)\boldsymbol{B}_j^{-1}(\boldsymbol{\beta}_j^\ast,\boldsymbol{\alpha}^\ast,\Lambda_{ij}^\ast)\boldsymbol{P}_j(\boldsymbol{\beta}_j,\boldsymbol{\alpha}^\ast,\Lambda_{ij}^\ast)\boldsymbol{\Omega}_j^{-1}(\boldsymbol{\gamma}_j^\ast)\boldsymbol{U}_{ij}(\boldsymbol{\gamma}_j^\ast) \\
    &+\boldsymbol{A}_j(\boldsymbol{\beta}_j^\ast)\boldsymbol{B}_j^{-1}(\boldsymbol{\beta}_j^\ast,\boldsymbol{\alpha}^\ast,\Lambda_{ij}^\ast)\boldsymbol{Q}_j(\boldsymbol{\beta}_j^\ast,\boldsymbol{\alpha}^\ast,\Lambda_{ij}^\ast)\int_0^{L_i}\frac{dM_{ij}^\ast(u)}{s^{(0)}(u;\boldsymbol{\gamma}_j^\ast)}.
\end{align*}
By the central limit theorem, $n^{-1/2}\sum_{i=1}^n\psi_{ij}$ converges in distribution to a normal distribution with mean $0$ and variance $E(\psi_{ij}^2)$.

\section{Methods under Comparison}
\label{s:comparative}
We compare our proposed method with several alternative approaches. The first is to leave out censored subjects and apply standard inverse probability weighted (IPW) method to the data with completely observed outcome only. The corresponding estimator for the average potential outcome in treatment group $j$ is given by
\begin{align*}
    \hat{\mu}_{j,\text{IPW}} = n^{-1}\sum_{i=1}^n\frac{D_{ij}R_i\tilde{Y}_i}{\pi_{ij}(\hat{\boldsymbol{\alpha}})}.
\end{align*}

The second approach considered builds on the IPW estimator and, along the line of Anstrom and Tsiatis,\cite{anstrom2001} further weights the subjects by the inverse probability of not being coarsened, namely, CIPW estimator
\begin{align*}
    \hat{\mu}_{j,\text{CIPW}} = n^{-1}\sum_{i=1}^n\frac{D_{ij}R_i\tilde{Y}_i}{\pi_{ij}(\hat{\boldsymbol{\alpha}})\exp\{-\hat{\Lambda}_{ij}(L_i)\}}.
\end{align*}

The third is the estimator of Wang et al,\cite{wang2016} which is a doubly robust estimator for average treatment effect using an augmented inverse probability weighted method, and we label it CAIPW-Wang. Let $h_{ij}(\boldsymbol{\omega}_j)$ be a posited model, in this case a logistic regression model, for $E(\tilde{Y}_i|Z_i=j, \tilde{\boldsymbol{X}}_i)$. The estimates for the parameter $\boldsymbol{\omega}_j$, denoted by $\hat{\boldsymbol{\omega}}_j$, are obtained by solving the score functions weighted by the inverse probability of not being censored. The final estimator is given by
\begin{align*}
    \hat{\mu}_{j,\text{CAIPW-Wang}} = \left(\sum_{i=1}^nw_i\right)^{-1}\sum_{i=1}^nw_i\left\{\frac{D_{ij}\tilde{Y}_i}{\pi_{ij}(\hat{\boldsymbol{\alpha}})}-\frac{D_{ij}-\pi_{ij}(\hat{\boldsymbol{\alpha}})}{\pi_{ij}(\hat{\boldsymbol{\alpha}})}h_{ij}(\hat{\boldsymbol{\omega}}_j)\right\},
\end{align*}
where $w_i=\sum_{i=1}^n\{\Delta_i/\sum_{j=1}^JD_{ij}\hat{K}_j[\min(T_i, C_i)]\}$ and $\hat{K}_j(t)$ is the treatment-specific Kaplan-Meier (KM) estimator.

The fourth is to apply standard causal inference methods, such as IPW, to pseudo-values of the outcome,\cite{andersen2017} and we call it Pseudo-IPW. Suppose that the parameter of interest is $\theta=E\{I(T_i\ge d)\}$. The pseudo-observations for subject $i$ is defined as $\theta_i=n\hat{\theta}-(n-1)\hat{\theta}^{-i}$, where $\hat{\theta}$ is the KM estimator and $\hat{\theta}^{-i}$ is the estimator applied to the sample from which subject $i$ is excluded. The method is implemented using the \textit{pseudo} package in R.\cite{perme2017} The pseudo-observations can be viewed as a replacement for the (possibly incompletely observed) outcome variable, and censoring is taken care of in the computation of pseudo-observations. In this case, the pseudo-observations are calculated assuming the independence of $T$ and $C$ given $Z$.

The fifth is the estimator of Zhang and Schaubel,\cite{zhang2012} which is originally designed for estimating the restricted mean lifetimes, and involves modeling the entire survival curve of the event time using Cox proportional hazards models, rather than focusing on a fixed time point as in the aforementioned approaches and our proposed method. It first estimates the cumulative hazard function for the event time $T$ for each group $j$, denoted by $\hat{\Lambda}_j(t)$, by augmenting an inverse probability weighted estimating equation with additional terms that involve outcome models. Then one can estimate the survival probability $\mu_j(t)$ at any time point $t$ by $\hat{\mu}_j(t)=e^{-\hat{\Lambda}_j(t)}$. Therefore, the method of Zhang and Schaubel can also be used for evaluating the risk at a specific time point $d$. We label their approach based on the inverse probability weighted estimating function as CIPW-ZS, and the augmented version as CAIPW-ZS. CAIPW-ZS is doubly robust under Assumption \ref{assump:censor}, such that the estimator is consistent when either the time-to-event or the coarsening mechanism is correctly modeled. 

The method of Zhang and Schaubel is developed within the general framework of augmented inverse probability weighting, whereas the proposed method is a standardization method. Other than this, another key difference between these two methods is that the former uses the Cox models but the proposed method uses logistic regression models  as working models to improve efficiency of the treatment effect estimator. If the interest only lies in a binary outcome (i.e., the occurrence of the event within a specific time point), theoretically one only needs to model the relationship of the binary outcome with covariates to improve efficiency. The method of Zhang and Schaubel requires modeling the relationship of the hazard function (equivalently, the survival curve), not limited to a specific time point,  with covariates. This tends to be an overkill for our purpose and increase the chance of model misspecification. When the Cox model is severely misspecified, CAIPW-ZS may provide little efficiency gain for the treatment effect estimates. We illustrate this point in one of our simulation settings.

We further consider the simple difference in the average outcome of each treatment group, as a benchmark, and call it the Naive estimator. The Naive estimator ignores both confounding and censoring. Sample code for the methods described in this section can be found at \url{https://github.com/youfeiyu/CIPWR}.

Among these methods, Naive and IPW estimators fail to account for censoring. Pseudo-IPW and AIPW-Wang assume that $T$ and $C$ are independent conditional on $Z$. CIPW-ZS and CAIPW-ZS, along with our proposed method CIPWR, rely on a more relaxed assumption that $T$ and $C$ are independent conditional on $Z$ and $\tilde{\boldsymbol{X}}$. CAIPW-Wang, CAIPW-ZS, and CIPWR leverage the information about the outcome model, which asymptotically improves the precision of the estimates. Moreover, the double robustness property of CAIPW-Wang and CAIPW-ZS are provided by the augmentation terms in the estimating equations, while the proposed CIPWR achieves double robustness through standardization of the weighted outcome model.

\section{Simulation Studies}
\label{s:simulation}
We compared the finite sample performance of our proposed method to the six alternative approaches described in Section \ref{s:comparative} through simulation studies. Specifically, we considered two settings that varied in degrees of nonproportionality with respect to the hazard functions for our simulation. The first setting assumed a logistic distribution for the time to event such that the hazard functions did not cross. The second setting concerned hazard functions that crossed at a certain time point for subjects with different covariate values, in which case Cox proportional hazards model tended to perform poorly in terms of improving precision.

\subsection{Simulation Setting I: Non-crossing Hazards}
\label{subs:setting1}
For the first setting, each simulated dataset contained five baseline covariates. $X_1$, $X_2$, and $X_3$ were independently sampled from a standard normal distribution. $X_4\sim\text{Bernoulli}(0.4)$ and $X_5\sim\text{Uniform}(-2,2)$. The treatment assignment $Z$ was simulated from a categorical distribution with the probability of receiving treatment $j$ being
\begin{align*}
    \frac{\exp(\alpha_{j0}+\alpha_{j1}X_1+\alpha_{j2}X_2+\alpha_{j4}X_4)}{\sum_{z=1}^3\exp(\alpha_{z0}+\alpha_{z1}X_1+\alpha_{z2}X_2+\alpha_{z4}X_4)}
\end{align*}
for $j=1,2,3$. The potential time to event $T^{(j)}$ was sampled from a logistic distribution with mean function $\beta_{j0} + \beta_{j1}X_1 + \beta_{j2}X_2+\beta_{j3}X_3$ and scale parameter $s=7$. The potential outcome $Y^{(j)}$ is defined as $Y^{(j)}=I\{T^{(j)}<d\}$, where $d=130$. We generated the censoring time $C$ using inverse transform sampling.\cite{Bender2005} In particular, we assumed a Cox proportional hazard model with the baseline hazard following a Weibull distribution,
\begin{align*}
    C^{(j)} = \{\lambda^{-1}\exp(\gamma_{j0}+\gamma_{j1}X_1+\gamma_{j2}X_2+\gamma_{j5}X_5)^{-1}\log u\}^{1/\nu},
\end{align*}
where the scale parameter $\lambda=0.01$, the shape parameter $\nu=7$, and $u$ was randomly sampled from a uniform distribution with interval [0,1].

We defined a `baseline' scenario where the outcome was weakly associated with the covariates and the proportion of being censored by $d=130$ was $30\%$. Then we varied the corresponding parameters to induce three proportions of being censored by $d=130$ ($20\%$, $30\%$, and $40\%$) and two levels of associations with the outcome (strong and weak). We also considered a scenario where censoring was independent of the covariates, referring to it as the random censoring scenario, with 30\% of the subjects being censored by $d=130$. The values of the parameters chosen in Setting I are listed in Table B.1. The true values for the estimands $E\{Y^{(1)}\}$, $E\{Y^{(2)}\}$, and $E\{Y^{(3)}\}$ were respectively 0.36, 0.50, and 0.63 for the scenarios of weak outcome associations, and 0.40, 0.59, and 0.50 when the outcome associations were strong.

The propensity scores were estimated using a multinomial logistic regression model, and the probability of remaining uncensored at $d$ was estimated by a Cox proportional hazards model. For the random censoring scenario, the probability of remaining uncensored was also estimated using the treatment-specific KM estimator. For the scenario with the largest proportion of censored observations ($\sim$40\%), we further considered the case in which the censoring time was observed for all subjects, as was the case in our data example, and evaluated the performance of CIPWR based on the observed censoring time (in contrast to the observation time). Across the scenarios, we considered three sets of model specifications for the CIPWR estimator: (1) correctly specified models for outcome, treatment, and censoring, (2) correctly specified models for treatment and censoring only, and (3) a correctly specified outcome model only. The misspecification for each model was caused by removing the confounder $X_2$. The CAIPW-ZS estimator assumed a Cox proportional hazard model for the survival time, and therefore the outcome model was always misspecified in this case. The CAIPW-Wang method only considered an outcome model and a propensity score model, since the survival function of the censoring time was estimated by the KM estimator. 

\subsection{Simulation Setting II: Crossing Hazards}
Two covariates independently sampled from the standard normal distribution, $X_1$ and $X_2$, were considered for the setting of crossed hazard functions. We assumed a multiphase model for the event time, where the effects of risk factors on the hazards differed by phases. The varying effects over time of risk factors are often seen in the setting of surgery. Specifically, the event time was generated such that the hazard functions crossed at some time point, and the equations we used to obtain the event time are listed in Section C of the Supplementary Materials. The probability of being assigned to treatment $j$ was $\exp(\alpha_{j1}X_1+\alpha_{j2}X_2)/\sum_{z=1}^3\exp(\alpha_{z1}X_1+\alpha_{z2}X_{2})$. Censoring was generated using $C=-\lambda^{-1}\exp\{\gamma_{1}X_1+\gamma_2X_2+\theta_1I(Z=2)+\theta_2I(Z=3)\}^{-1}\log u$. In the first scenario, we assumed that the treatment assignment and censoring time only depended on $X_1$, such that $\boldsymbol{\alpha}_1=(\alpha_{11},\alpha_{12})^T=(0,0)^T$, $\boldsymbol{\alpha}_2=(\alpha_{21},\alpha_{22})^T=(0.2,0)^T$,
$\boldsymbol{\alpha}_3=(\alpha_{31},\alpha_{32})^T=(0.3,0)^T$, and $(\lambda,\gamma_1,\gamma_2,\theta_1,\theta_2)^T=(0.8,1,0,0.2,0.4)^T$. In the second scenario, we let $X_2$ come into play, such that $\boldsymbol{\alpha}_1=(\alpha_{11},\alpha_{12})^T=(0,0)^T$, $\boldsymbol{\alpha}_2=(\alpha_{21},\alpha_{22})^T=(0.2,0.2)^T$,
$\boldsymbol{\alpha}_3=(\alpha_{31},\alpha_{32})^T=(0.3,0.3)^T$, and $(\lambda,\gamma_1,\gamma_2,\theta_1,\theta_2)^T=(0.7,-0.5,0.5,0.4,0.2)^T$. The cutoff point $d$ was chosen to be 0.5 and 0.3 for the first and second scenario, respectively, which led to 30.7\% and 13.2\% of censored observations for the corresponding $d$. The true values for $E\{Y^{(1)}\}$, $E\{Y^{(2)}\}$, and $E\{Y^{(3)}\}$ were 0.68, 0.62, and 0.52 for the first scenario, and 0.54, 0.45, and 0.41 for the second scenario.

The models for the treatment assignment and censoring were correctly specified in this setting. The logistic regression model and the Cox model for the outcome were misspecified such that squared terms of the covariates and interactions (if applicable) were included in the model.

\subsection{Evaluation Metrics}
For each scenario, we generated $2000$ Monte Carlo datasets, each with $n=1500$ subjects. For CIPWR, the standard errors and 95\% confidence intervals (CIs) were estimated using the formula for asymptotic variance derived in Section \ref{s:asymptotic}. For the other methods considered, 200 bootstrap replicates were used to estimate the standard errors and 95\% CI. Simulation results are presented in terms of bias, empirical standard deviation, root mean squared error (RMSE) and coverage rate of 95\% CI.

\subsection{Simulation Results}
The numerical results for the simulation are reported in Tables B.2-B.7 in the Supplementary Materials, and are summarized in Figures 1-3 in the main text and Figures B.1-B.7 in the Supplemental Materials. Figures B.1-B.5 show the results for bias for the methods under comparison. Naive and IPW estimates were expectedly biased away from the true risk differences in all the scenarios in both settings (Figures B.1-B.3 in the Supplemental Materials), as they failed to accommodate censoring. When censoring was unrelated to baseline covariates (the random censoring scenario), the rest of the estimators considered were consistent for the true values (Figure B.1 in the Supplemental Materials). CIPWR estimators had close to zero bias regardless of whether the probability of remaining uncensored was estimated by Cox model or KM estimator (Figures B.4 in the Supplemental Materials). When censoring depended on the covariates, the bias for Pseudo-IPW and CAIPW-Wang estimators, which relied on the restrictive independence assumption $C\condind T|Z$, became non-negligible. For example, the relative bias for Pseudo-IPW with a correctly specified propensity score model reached 21.5\% in the scenario with 40\% censoring (Table B.5 in the Supplemental Materials). CAIPW-ZS and CIPWR with at least one of the models for outcome and coarsening mechanism correctly specified remained unbiased across the scenarios considered in both settings, exhibiting double robustness property (Figures B.1-B.3 in the Supplemental Materials). When the probability of remaining uncensored was estimated based on the observed censoring time rather than the observation time, the empirical bias for CIPWR remained close to zero (Figure B.5 in the Supplemental Materials). 

\begin{figure}[t]
 \centerline{\includegraphics[width=\textwidth]{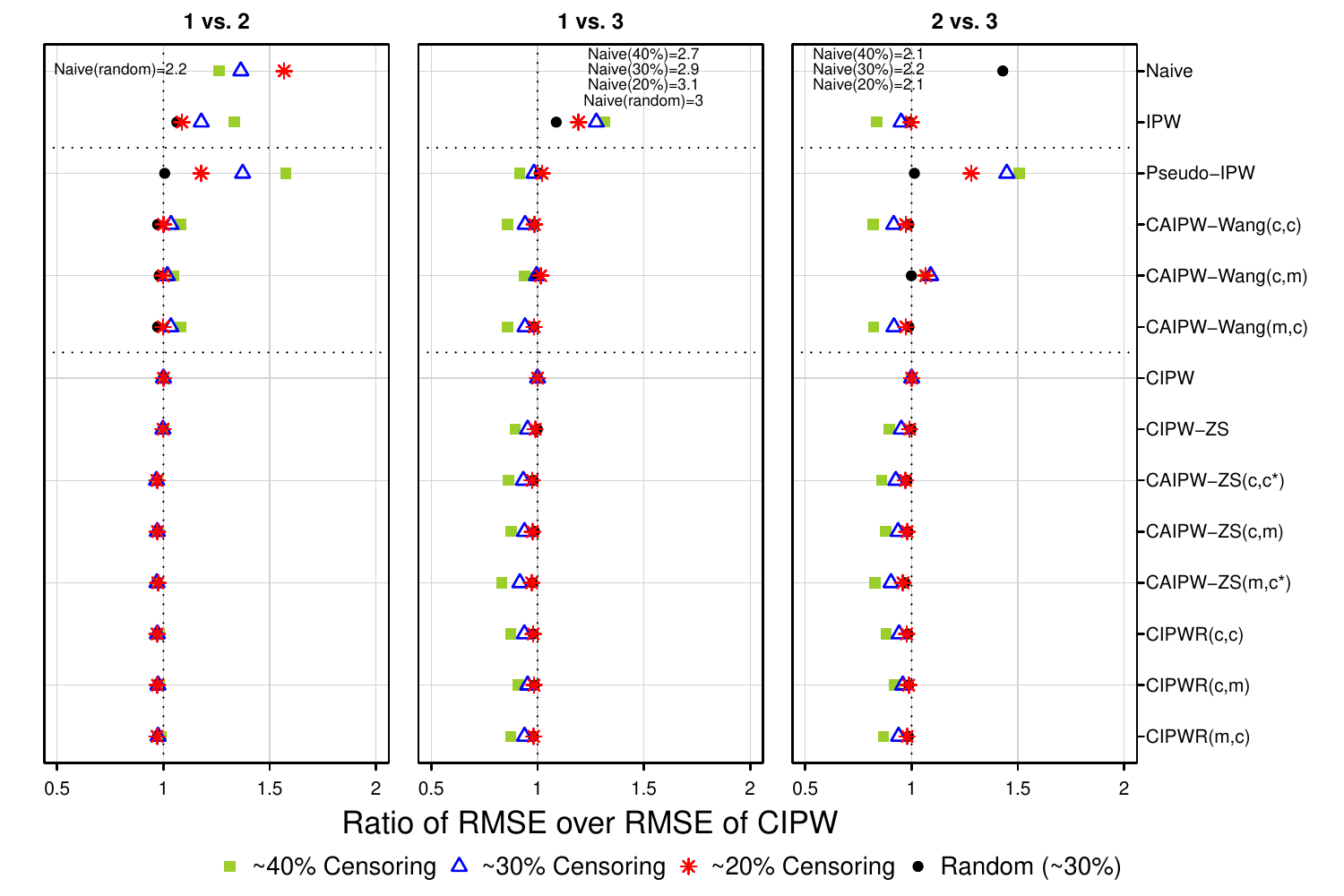}}
\caption{RMSE over RMSE of CIPW with correctly specified propensity and censoring models for different proportions of censoring in Setting I. For CAIPW-Wang, the first letter and second letter denote the specification of the propensity and outcome model, respectively. For CIPWR and CAIPW-ZS, the first and second letter in the parentheses correspond to the model for coarsening mechanism and outcome, respectively. The outcome model in CAIPW-ZS is always misspecified, and we use c$^\ast$ to denote the case where the true predictors for the outcome were included in the model. Propensity model is correctly specified for IPW, Pseudo-IPW, CIPW, and CIPW-ZS. Numbers that fall outside the range of x-axis are labeled in the figure. Sample size was 1500. Results were obtained using 2000 simulated datasets.}
\label{fig:RMSE_censor}
\end{figure}

\begin{figure}[t]
 \centerline{\includegraphics[width=\textwidth]{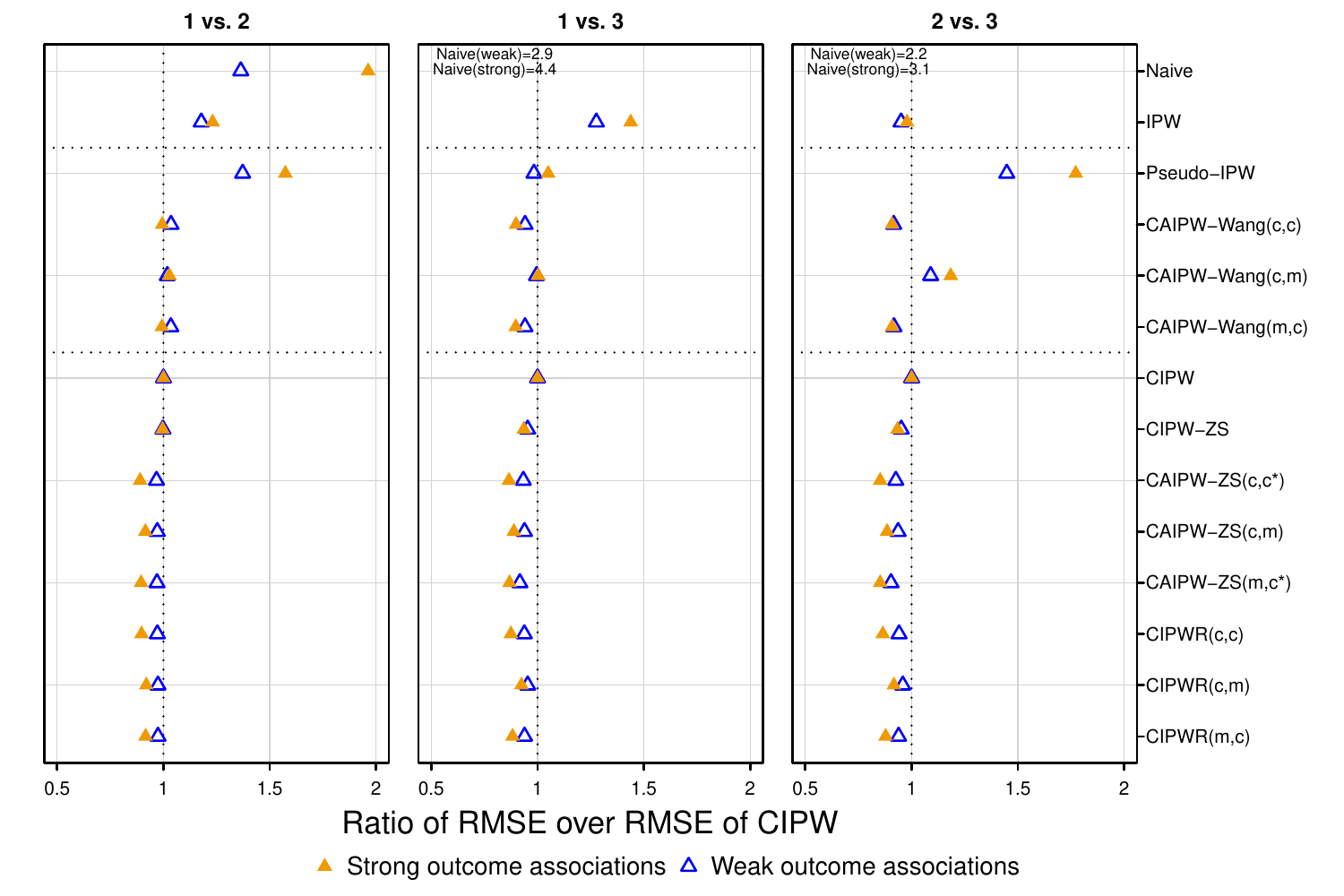}}
\caption{RMSE over RMSE of CIPW with correctly specified propensity and censoring models for different levels of outcome associations in Setting I. Censoring depended on covariates and censoring proportion was 30\%. For CAIPW-Wang, the first letter and second letter denote the specification of the propensity and outcome model, respectively. For CIPWR and CAIPW-ZS, the first and second letter in the parentheses correspond to the model for coarsening mechanism and outcome, respectively. The outcome model in CAIPW-ZS is always misspecified, and we use c$^\ast$ to denote the case where the true predictors for the outcome were included in the model. Propensity model is correctly specified for IPW, Pseudo-IPW, CIPW, and CIPW-ZS. Numbers that fall outside the range of x-axis are labeled in the figure. Sample size was 1500. Results were obtained using 2000 simulated datasets.}
\label{fig:RMSE_outcome}
\end{figure}

\begin{figure}[t]
 \centerline{\includegraphics[width=\textwidth]{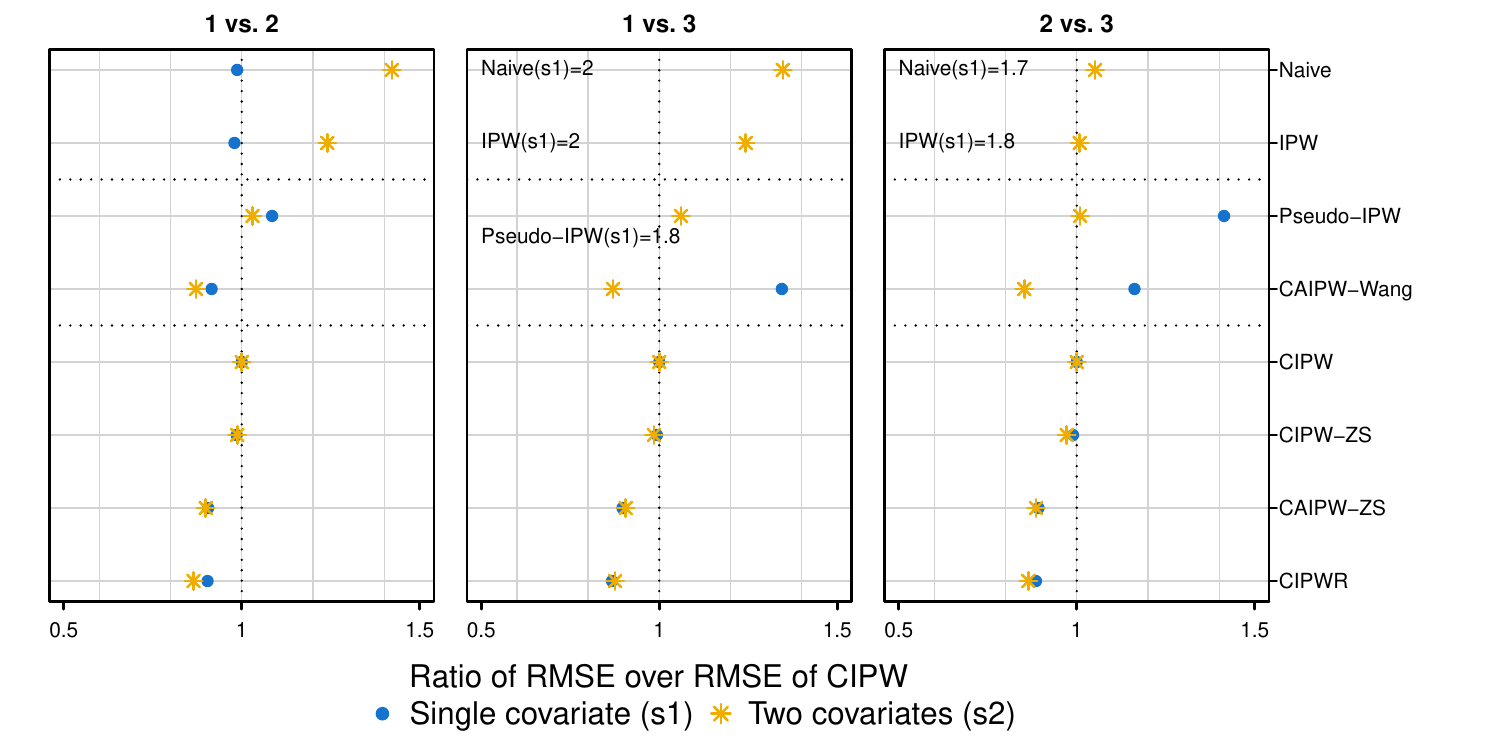}}
\caption{RMSE over RMSE of CIPW with correctly specified propensity and censoring models in the presence of crossing hazards in Setting II. The models for the coarsening mechanism were correctly specified. The outcome model was always misspecified in this setting. Numbers that fall outside the range of x-axis are labeled in the figure. Sample size was 1500. Results were obtained using 2000 simulated datasets.}
\label{fig:RMSE_nonPH}
\end{figure}

The RMSE results for Setting I are displayed in Figures \ref{fig:RMSE_censor} and \ref{fig:RMSE_outcome}. We report the ratio of RMSE to the RMSE of CIPW with correctly modeled coarsening mechanism. Figure \ref{fig:RMSE_censor} summarizes the results for different censoring mechanisms and proportions of being censored. In general, CAIPW-ZS and CIPWR had the smallest RMSE among the methods considered across all treatment pair comparisons. As the censoring proportions increased, we observed larger gain in efficiency for CAIPW-ZS and CIPWR over CIPW. Cox model-based and KM estimator-based CIPWR yielded similar RMSE in the random censoring scenario (Figure B.6 in the Supplemental Materials), with the former having slightly smaller RMSE than the latter in some cases. This finding suggests that estimating the probability of remaining uncensored from the data, even if the value is known, may actually lead to smaller variance for the CIPWR estimator than using the true value, which is consistent with the theoretical results in the literature.\cite{tsiatis2006} Moreover, estimating the probability of remaining uncensored using observation time tended to reduce the RMSE compared with using observed censoring time (Figure B.7 in the Supplemental Materials). Figure \ref{fig:RMSE_outcome} displays the RMSE results for different levels of outcome associations. When the associations between the covariates and the outcome were weak, we observed 2.6\%-3.7\% reduction in RMSE for CIPWR and CAIPW-ZS. Greater reduction (7.6\%-10.5\%) was noted as the associations became stronger. The RMSE results for Setting II where crossing hazards existed are presented in Figure \ref{fig:RMSE_nonPH}. Note that in this setting both the logistic regression model and the Cox model for the outcome were misspecified. Again, we observed lower RMSE for CIPWR and CAIPW-ZS compared to CIPW. Furthermore, CAIPW-ZS produced larger variability (and therefore larger RMSE) than CIPWR when the hazard functions had a crossing (Figure~\ref{fig:RMSE_nonPH} and Table B.7). For example, the ratios of the variance of CAIPW-ZS to that of CIPWR ranged from 1 to 1.06 in the first scenario, and from 1.05 to 1.08 in the second scenario. 

All methods that were approximately unbiased in the presence of nonrandom censoring (i.e., CIPW, CIPW-ZS, CAIPW-ZS, and CIPWR) achieved close to nominal coverage of 95\% (Tables B.2-B.7 in the Supplemental Materials).

\section{Application to Comparison of Treatments for Prostate Cancer using Medical Claims}
\label{s:application}
\subsection{Data Analysis Methods}
We applied our proposed method to a data set comprised of patients with metastatic castration-resistant prostate cancer (mCRPC), which was obtained from a large national private health insurance network (Optum Clinformatic Data Mart). The study cohort included patients who used at least one of the six drugs (docetaxel, abiraterone, enzalutamide, sipuleucel-T, cabazitaxel, and radium-223) approved to treat mCRPC from January 1, 2014, to December 31, 2019. Among these drugs, docetaxel and cabazitaxel are chemotherapies, abiraterone and enzalutamide are oral hormone therapies, sipuleucel-T is an immunotherapy, and radium-223 is a radioactive drug. We excluded the patients who received cabazitaxel ($n=56$) or radium-223 ($n=28$) as their first-line therapy from our analysis, since there were much fewer samples in these two groups than the other four. We examined the occurrence of ER visits and all-cause hospitalization within 180, 270, and 360 days of treatment initiation, respectively. Patients who switched to another treatment or dropped out of the insurance plan prior to the event of interest within the pre-specified time window were considered as being censored.

The treatment was modeled using a multinomial logistic regression adjusting for age, race, education level, household income, geographic region, insurance product type, whether the insurance plan is administrative services only (ASO), metastatic status of cancer, year of first prescription, comorbid conditions, and provider type.\cite{caram2019} All covariates were binary or categorical. To improve the common support of the covariate distributions, we followed the criteria discussed in Lopez and Gutman\cite{lopez2017} and discarded the tails of the propensity score distributions. The outcome model and the censoring model adjusted for the same set of covariates that were controlled for in the treatment model. The CIs were obtained using (1) Wald-type CIs based on original data for the Naive method, and (2) bootstrap standard errors based on 200 bootstrap samples for the rest of the methods.

\subsection{Data Analysis Results}
Patients who had less than 180 days of continuous enrollment prior to the first prescription, had missing covariates, or experienced the event of interest on the same day as the first prescription were removed from the analysis. In the end, we identified 7678 and 7709 mCRPC patients for ER visit and hospitalization, respectively, and calling them ER visit cohort and hospitalization cohort. The sample sizes of the two cohorts differed because ER visit and initial treatment prescription for the first time were more likely to occur on the same day than subsequent hospitalization. The proportions of overall and cause-specific censoring within each specified time window for the two outcomes are reported in Table B.8 in the Supplementary Materials. In general, hospitalization (24.6\%-40.6\%) was associated with greater overall proportion of censored observations than ER visits (20.8\%-32.6\%). The proportions of patients being censored and the unadjusted risks ignoring censored patients for each treatment group are presented in Table B.9 in the Supplementary Materials. In particular, Sipuleucel-T group had larger percentage of censored patients than the other three groups. Docetaxel group had the highest crude risks of ER visits (53.6\%, 64.6\%, and 71.8\% within 180, 270, and 360-day time windows, respectively) and hospitalization (41.1\%, 52.3\%, and 60.2\% within 180, 270, and 360-day time windows, respectively). The baseline demographic and clinical characteristics of the ER visit cohort stratified by treatment groups are presented in Table B.10 in the Supplemental Materials. The covariate distributions of the hospitalization cohort were close to those of the ER visit cohort (data not shown). 

To improve the covariate overlap among the treatment groups for the comparative analysis, we applied data trimming with criteria discussed in Lopez and Gutman,\cite{lopez2017} which left us with 7003 and 7045 patients for ER visit and hospitalization, respectively.

\begin{figure}[t]
  \centerline{\includegraphics[width=\textwidth]{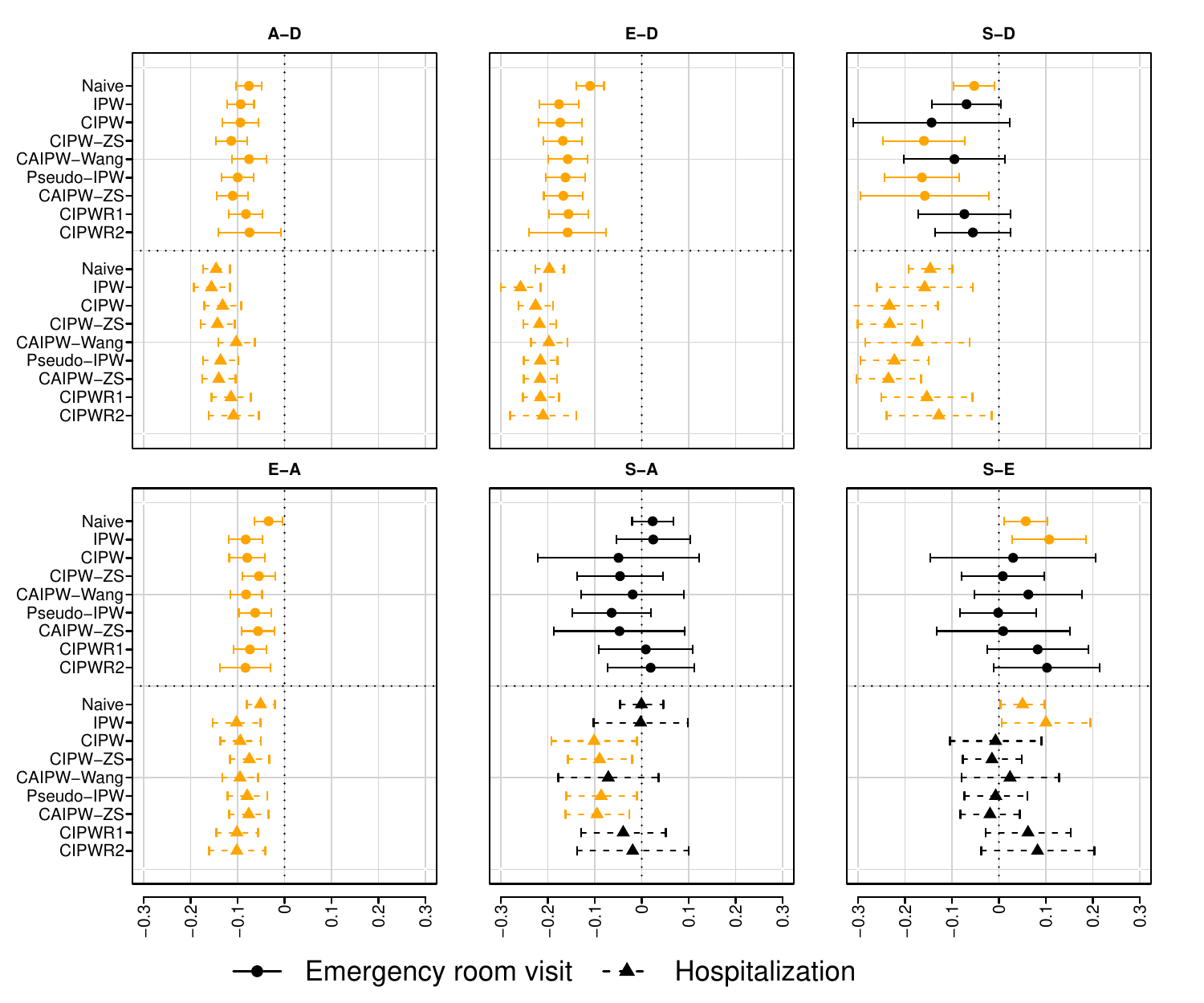}}
  
  \caption[Differences in 360-day risks of experiencing at least one emergency room visit or hospitalization among the four focus drugs and the associated 95\% confidence intervals.]{Differences in 360-day risks of experiencing at least one emergency room visit or hospitalization among the four focus drugs and the associated 95\% confidence intervals. Data were obtained from Optum Clinformative Data Mart. Total sample size was $N=7003$ ($N_A=2458$, $N_D=2162$, $N_E=1833$, $N_S=550$) for ER visits, and $N=7045$ ($N_A=2474$, $N_D=2172$, $N_E=1843$, $N_S=556$) for hospitalization. CIPWR1 was based on observation time, and CIPWR2 was based on observed censoring time. Confidence intervals that exclude zero are highlighted in orange. Abbreviations: A, abiraterone; D, docetaxel; E, enzalutamide; S, sipuleucel-T; ER, emergency room visit.}
\label{fig:analysis_d360}
\end{figure}

Figure~\ref{fig:analysis_d360} shows the differences in 360-day risks among the four treatment groups for both outcomes of interest. Results for 180-day and 270-day risks are presented in Figures B.8 and B.9 in the Supplemental Materials, respectively. When confounding and censoring were both ignored, docetaxel users had significantly higher risk of at least one ER visit within 360 days of treatment initiation than users of abiraterone, enzalutamide, and sipuleucel-T. Similar directional results for docetaxel vs. abiraterone and docetaxel vs. enzalutamide comparisons were noted for methods that accounts for both confounding and censoring, and the corresponding 95\% CIs consistently excluded zero across the methods. For example, the risk difference estimated by the CIPWR estimator using observation time (CIPWR1) was -0.082 (95\% CI [-0.118, -0.046]) for docetaxel vs. abiraterone comparison, and -0.156 (95\% CI [-0.198, -0.114]) for docetaxel vs. enzalutamide comparison. These findings agree with the clinical evidence that oral therapies abiraterone and enzalutamide tend to have fewer side effects than docetaxel, a chemotherapy.\citep{tonyali2017} Sipeucel-T was identified to have lower risk of ER visit than docetaxel, though the differences were not significant for some of the methods (e.g., risk difference=-0.158, 95\% CI [-0.295, -0.021] for CAIPW-ZS; risk difference=-0.073, 95\% CI [-0.172, 0.025] for CIPWR1). For the two oral drugs, Enzalutamide was identified to have lower risk of ER visit than Abiraterone within each specified specified period of time when both confounding and censoring were accounted for. The Naive and IPW methods indicated significantly higher 360-day risk of ER visit for Enzalutamide than Siputeucel-T, while the methods that account for both confounding and censoring showed that there was no significant difference. Similar patterns were observed for the risks of all-cause hospitalization for each time window considered.

In general, CIPWR based on observed censoring time (CIPWR2) tended to have wider CIs than CIPWR using observation time (CIPWR1), which is consistent with our simulation results (Figure B.7 in the Supplementary Materials). For example, the ratios of confidence widths of CIPWR1 over CIPWR2 for 360-day hospitalization ranged from 54.9\% to 86.9\%. Greater differences in the width of CIs were noted as the duration of time window became longer and the proportion of censored patients increased (Figures B.8 and B.9 in the Supplementary Materials and Figure \ref{fig:analysis_d360}). In most cases, CAIPW-ZS and CIPWR yielded narrower CIs than the CIPW estimator. For example, the percentage of reduction in width ranged from 58.8\% to 92.4\% for the risk of ER visits estimated by CIPWR1. CAIPW-ZS was noted to have wider CIs than CIPWR for treatment pairs that involved Sipuleucel-T for ER visits (Figure \ref{fig:analysis_d360}). Greater differences in point estimates between CAIPW-ZS and CIPWR, the former of which modeled the entire survival curve over time, was noted as the ending time point moved farther away from the treatment initiation. 

Differences between methods that ignore and account for censoring increased as the time window was extended. Results for methods that rely on different independence assumptions on the censoring mechanism were similar. One possible reason is that censoring may only depend on the treatment in this cohort, and the restrictive version of the assumption is satisfied, as the censored patients within each time window exhibited similar demographic and baseline clinical characteristics to uncensored ones (Table B.11 in the Supplemental Materials), and most covariates were not significantly associated with censoring (Table B.12 in the Supplemental Materials).

\section{Discussion}
\label{s:discussion}

We present an inverse probability weighted regression-based estimator, CIPWR, for average treatment effect for a binary outcome that is subject to right-censoring. This method is based on the intuitively simple standardization idea, where we model the binary outcome given the observed covariates using the familiar logistic regression model for each treatment separately and then averaging predictions for all patients. The CIPWR method improves robustness by accounting for confounding due to nonrandomized treatment and censoring using the inverse probability weighting approach. Therefore, the proposed method is a hybrid of the two general approaches (standardization and weighting) in the missing data and causal inference literature that handle missingness, confounding and censoring. Like the well-studied augmented inverse probability weighting approach (e.g., CAIPW-Wang and CAIPW-ZS), the proposed method enjoys a double robustness property such that the estimator is consistent if either the (binary) outcome, or both treatment assignment and censoring are correctly modeled. However, in this method the  double robustness and improvement in efficiency  are not through direct augmentation. Instead, it achieves double robustness by combining two approaches in different stpdf, with each step based on popular models and the natural ideas of standardization and weighting. The proposed method is conceptually straightforward to understand and easy to implement using standard statistical software for practitioners. 

Simulation studies show that in finite sample, CIPWR yielded approximately unbiased estimates and close to nominal coverage of 95\% across the scenarios considered, particularly when censoring depends on the covariates. CIPWR also provides efficiency gain over CIPW by exploiting the information from the outcome model. The proposed method was applied to claims data for comparing the average treatment effects of multiple treatments.

Time-to-event data are often analyzed using approaches that model the whole survival curve from baseline to the end of follow-up, such as the Cox proportional hazards model used for CAIPW-ZS.\cite{zhang2012} In the case where interest only lies in the risk difference over a pre-specified period of time (e.g., 180-day risk of ER visit), a method that directly targets the survival function at the fixed time point can lead to better efficiency than general methods that estimate the whole survival curve. The problem can be reduced to estimating the marginal expectation of a (possibly censored) binary indicator of event occurrence, which we propose to solve by utilizing logistic regression that directly targets the binary outcome. In general, the difficulty of correctly modeling the time-to-event outcome given observed covariates increases as the time window becomes longer, and the incorrect model for the survival time tends to result in more severe problems than misspecification of the outcome model at a fixed time point. In our simulation setting where the hazard functions did not cross, CIPWR, which directly modeled the binary indicator of event occurrence, performed similarly to CAIPW-ZS, which utilized outcome information accumulated over time, in terms of RMSE. In the presence of crossing hazards, when both the logistic regression model and Cox model were misspecified, CIPWR realized more efficiency gain over CIPW than CAIPW-ZS, as the latter failed to capture the associations between the outcome and covariates. The efficiency gain of CIPWR over CAIPW-ZS was also observed in the data example, where CAIPW-ZS had wider CI than CIPWR for some treatment pairs (Figure~\ref{fig:analysis_d360}), and the difference in confidence width increased as the time window became larger (Figure \ref{fig:analysis_d360} and Figure B.8 in the Supplementary Materials). 

In the presence of right censoring, discarding the censored observations may result in biased estimates for the treatment effects, even if the censoring was completely random, as shown in our simulation studies (Figure B.1 in the Supplementary Materials). In this paper, we assumed the independence of survival and censoring time conditional on treatment and baseline covariates, and used Cox model to estimate the probability of remaining uncensored, which was then inverted and used as weights in the estimating equation. Under the more restrictive conditional independence assumption given treatment only, it is sufficient to use the non-parametric KM estimator for estimating the probability of remaining uncensored. However, simulation results showed that when censoring was random, CIPWR based on Cox model was still unbiased for the treatment effect, and could possibly be more efficient than CIPWR based on KM estimator. 

In this study, we focus on low-dimensional covariates and only consider simple parametric working models for the outcome and treatment. As researchers gain increasing access to large databases with a substantial collection of covariates, variable selection techniques for causal inference has been an emerging topic of interest.\cite{shortreed2017,athey2018} Another possible extension to our method is to replace the parametric models with modern machine learning methods that can capture potential nonlinearities and nonadditivities. For example, neural networks and methods based on recursive partitioning have been suggested as promising alternatives to logistic regression for estimating propensity scores when the true model structure is complex.\cite{setoguchi2008,lee2010} In addition, treatment switching was treated in the same way as dropout and study termination. That is, an observation was considered censored when someone switched treatment, which may not be optimal and studying treatment sequences will be another challenge.


\section*{Acknowledgments}
The research of BM was supported by NSF DMS 1712933.

\subsection*{Conflict of interest}
The authors declare no potential conflict of interests.

\section*{Data Availability Statement}
The data that support the findings of this study are available on request from the corresponding author. The data are not publicly available due to privacy or ethical restrictions.

\section*{Supporting information}

Additional supporting information may be found online in the Supporting Information section at the end of this article.

\bibliography{CIPWR_refs}%

\clearpage

\end{document}